\documentclass[11pt,twoside]{article}

\usepackage{asp2006}
\usepackage{epsf}
\usepackage{psfig}
\usepackage{lscape}

\begin{document}
\title{The role of the star formation on the nitrogen abundance evolution}  

\author{M. Moll\'{a},\altaffilmark{1}, J. M. V\'{\i}lchez,\altaffilmark{2}
 A.I. D\'{\i}az,\altaffilmark{3} and M. Gavil\'{a}n.\altaffilmark{3}}

\altaffiltext{1}{CIEMAT,Departamento de Investigaci\'{o}n B\'{a}sica, 
Avda. Complutense-22, 28040 Madrid, (Spain)}

\altaffiltext{2}{Instituto de Astrof\'{\i}sica de Andaluc\'{\i}a, CSIC, 
Apdo. 3004, 18080 Granada, (Spain)}

\altaffiltext{3}{Departamento de F\'{\i}sica Te\'{o}rica,
Universidad Aut\'onoma de Madrid, 28049 Cantoblanco, Madrid 
(Spain)}

\begin{abstract} 
We analyze the evolution of nitrogen resulting from a set of
spiral and irregular galaxy models computed for a large number of input
mass radial distributions and with various star formation efficiencies.
We show that our models produce a nitrogen abundance evolution in good
agreement with the observational data. Differences in the star 
formation histories of the regions and galaxies modeled are essential to
reproduce the observational data in the N/O-O/H plane and the
corresponding dispersion.
\end{abstract}

\section{The primary Nitrogen} 

When observations, specially the metal-poor galaxy data, are
ploted in the plane N/O {\sl vs} O/H, it is evident that a 
primary N contribution must exist. Applying the classical 
{\sl Closed Box Model}, the points may be limited by three lines as those
plotted in Fig.~\ref{no_hii}a:
1) This one called NS, which shows the evolutionary tracks of N/O when 
N need a seed of O to be created: 
$\frac{N}{O}=\frac{p_{N}(O)}{p_{O}}\propto O $.
2) This one called NP that appear when N is created directly from H or He:
$\frac{N}{O}=\frac{p_{N}}{p_{O}}= constant$. 
3) NS$+$NP when both contributions there exist, which seems to be the
apparent behavior reproduced by the data.

It is, however, necessary to take into account three factors not
included in that simple scenario: 1) the mean lifetimes of stars, 2)
the star formation efficiency or the existence of different star
formation histories or star formation rates in different galaxies or
objects, and 3) the metallicity dependent of stellar yields.  If the
NP is ejected by low and intermediate mass (LIM) stars, it will appear in the
interstellar medium very abruptely after a time delay, which means
when O/H has already reached a certain value.  Obviously, this level
will be higher or lower depending on the mean-lifetime of these stars. But it
also depends on the efficiency to form stars: for a high efficiency
the NP will appear later in the time, or at higher O/H, than for a low
one. Both facts produce a certain dispersion in the resulting
abundances. Besides that the metallicity dependent stellar yields
produce tracks elongated in comparison with the one produced when only
a value of yield is used \cite[for details see][]{mol06}.

\section{The multiphase chemical evolution models grid.}

We have computed a grid of models \citep{mol05} with 44 theoretical
galaxies of different total masses and 10 possible efficiences to form
stars in each one. We have used the yields from \cite{ww95} for
massive stars and those from \cite{gav05} for LIM
stars, which give results for our Galaxy \citep{gav06}
in excelent agrement with halo stars data \citep{isr04,spi05}.

\begin{figure}[!ht]
\epsfclipon 
\plottwo{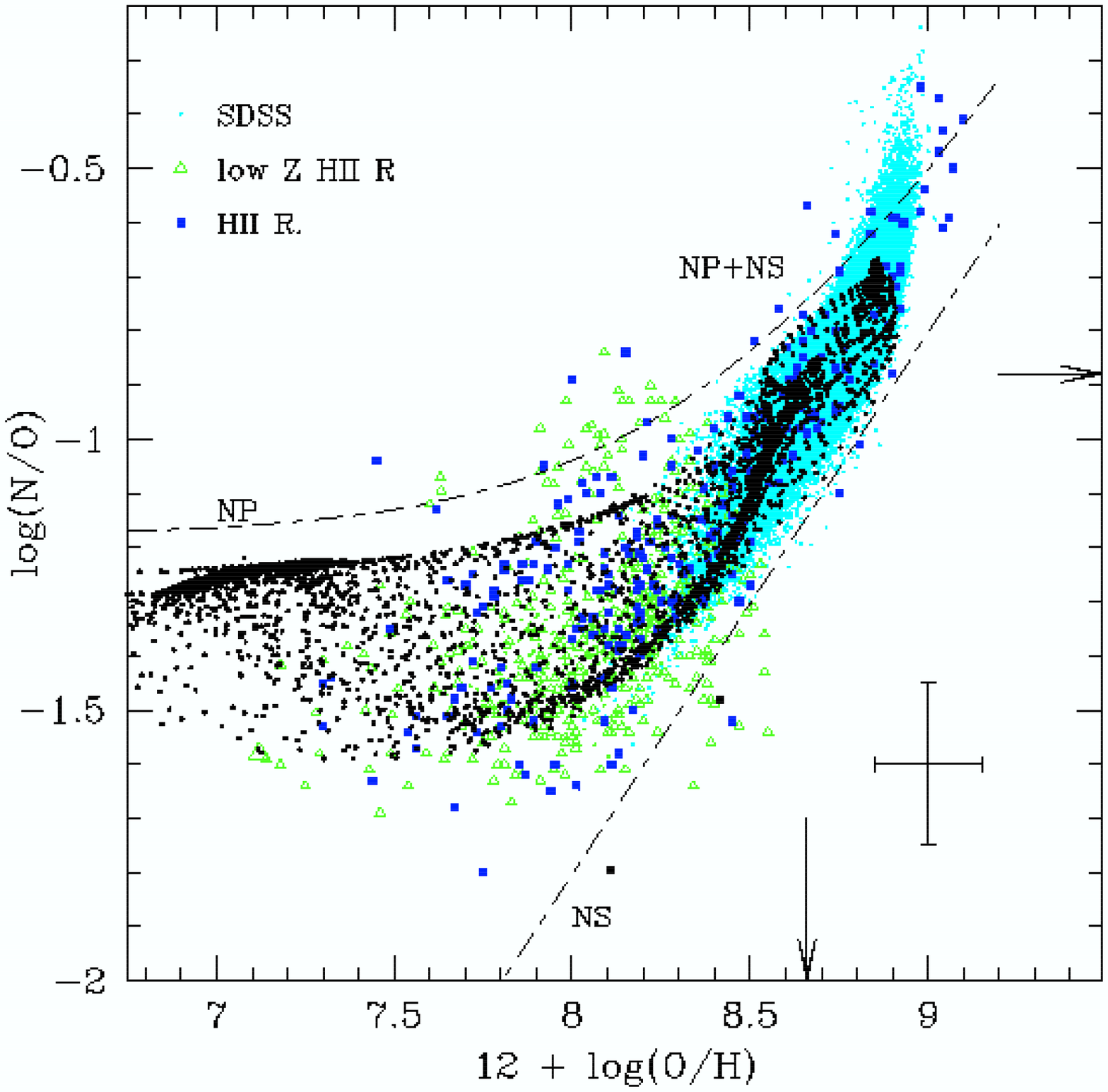}{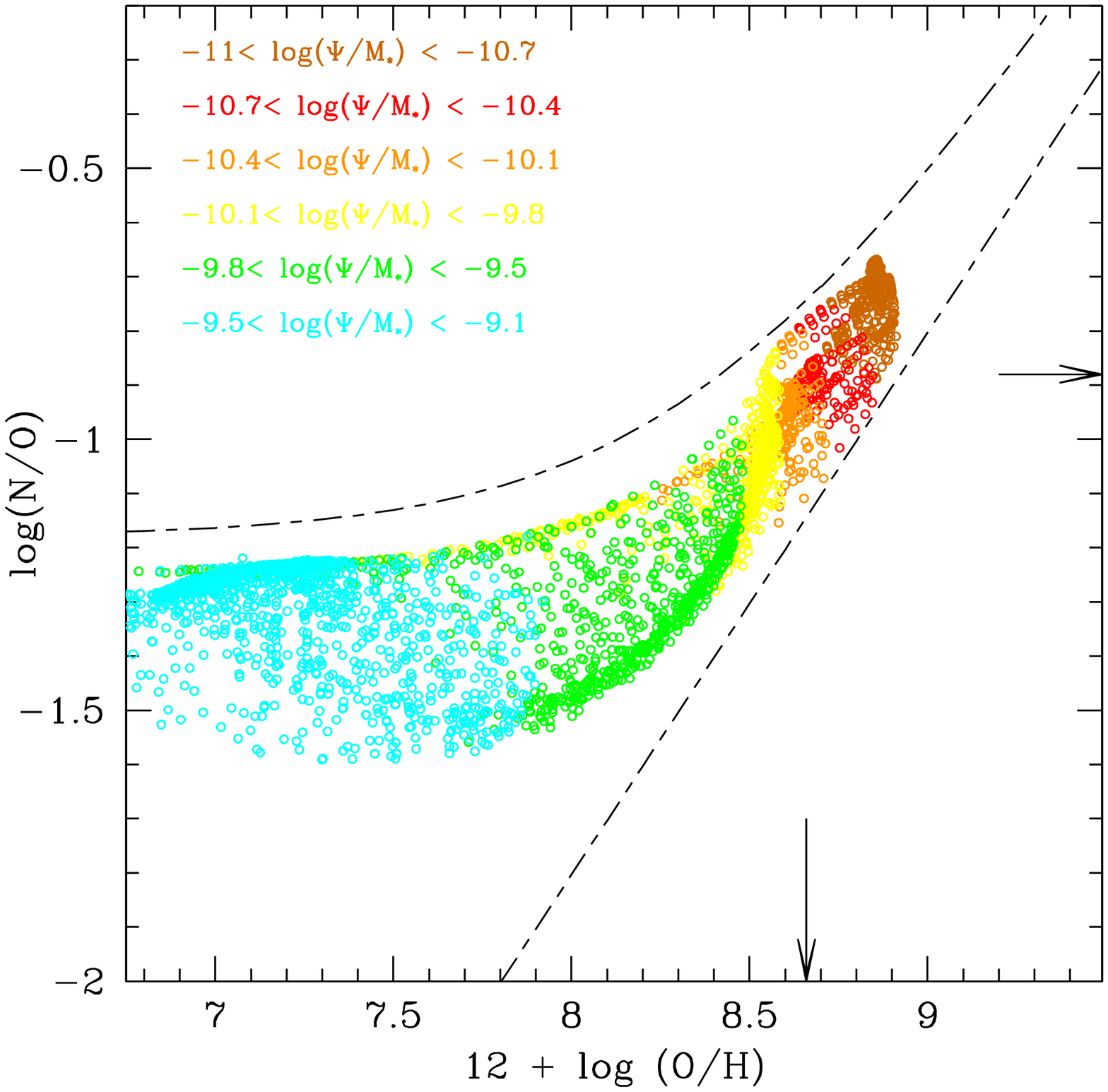}
\caption{The relative abundance $log(N/O)$ {\sl vs} the oxygen
abundance $12 + log(O/H)$: a) Our model results for the present
time as small black dots compared with data (colors as labelled)
from authors given in Table~1 of \cite{mol06}; b) The same results are represented
according their present specific star formation rate as labelled.}
\label{no_hii}
\end{figure}

The obtained results \citep{mol06} for the present time are compared
with observations in Fig.~\ref{no_hii}a). The position of a galaxy in
the plane N/O-O/H depend on the star formation histories and/or on the
actual star formation rate as we may see in Fig.~\ref{no_hii}b): if
the star formation occured as a strong and early burst, with a low
rate at the present time compared with the past maximum (red dots),
the evolutionary track in that plane appears as very secondary and O/H
and N/O are high. When the star formation occurs quietly with a rate
higher now than in the past (cyan dots), the track is almost flat with
low O/H and N/O abundances. This way we conclude that our grid of
models reproduce and explain very well most of observational data in
the plane N/O-O/H.

\end{document}